\documentclass[12pt]{iopart}
\usepackage{iopams,graphicx}

\begin{document}
\title[Polarization of the vacuum by an impenetrable magnetic vortex]{Polarization of the vacuum of a quantized scalar field by
an impenetrable magnetic vortex of finite thickness}

\author{V M Gorkavenko$^1$, Yu A Sitenko$^2$ and O B Stepanov$^1$}

\address{$^{1}$ Department of Physics, Taras
Shevchenko National University of Kyiv, 64 Volodymyrs'ka str., Kyiv
01601, Ukraine}
\address{$^2$ Bogolyubov Institute for Theoretical Physics,
 National Academy of Sciences of Ukraine, 14-b Metrologichna str.,
Kyiv 03680, Ukraine} \ead{\mailto{gorka@univ.kiev.ua}}

\begin{abstract}
We consider the effect of the magnetic field background in the form
of a tube of the finite transverse size on the vacuum of the
quantized charged massive scalar field which is subject to the
Dirichlet boundary condition at the tube. It is shown that, if the
Compton wavelength associated with the scalar field exceeds
considerably the transverse size of the tube, then the vacuum energy
which is finite and periodic in the value of the magnetic flux
enclosed in the tube is induced on a plane transverse to the tube.
Some consequences for generic features of the vacuum polarization in
the cosmic-string background are discussed.
\end{abstract}
\pacs{11.27.+d, 11.10.Kk, 11.15.Tk} \submitto{\JPA} \maketitle

\normalsize

\section{Introduction}

The emergence of calculable and detectable vacuum energy as a
consequence of imposing external boundary conditions in quantum
field theory was predicted more than $60$ years ago by Casimir
\cite{Cas}. Since that time the vacuum energy of fluctuating quantum
fields that are subject to boundary conditions was studied in
various setups (see, e.g., reviews in \cite{Eli,Most,Bordag}).
Usually, the boundary manifold is chosen to be noncompact
disconnected  (e.g., two parallel infinite plates, as generically in
\cite{Cas}) or closed compact (e.g., box or sphere); see
\cite{Eli,Most,Bordag}.

In the present paper, we shall consider  the boundary manifold which
is noncompact connected and has the form of an infinite tube in
three-dimensional space. As has been first demonstrated by Aharonov
and Bohm \cite{Aha} in the framework of first-quantized theory, the
magnetic flux enclosed in such a tube affects the properties of
quantum matter outside the tube. This effect which is named after
them has no analogues in classical physics and is characterized by
the periodic dependence on the value of the flux, vanishing at
integer multiples of the London flux value, while being maximal at
the half of the London flux value.

In the framework of second-quantized theory, one is interested in
the vacuum po\-la\-ri\-zation effects which are induced outside the
tube by the magnetic flux enclosed in the tube. In particular, the
question is, whether the vacuum energy is induced. If this is the
case, then this may be denoted as the Casimir-Bohm-Aharonov effect
(see also \cite{Sit}).

It should be noted that initially \cite{Aha}  the Bohm-Aharonov
effect was considered under the assumption that the transverse size
of the tube is zero, which corresponds to the singular magnetic
vortex configuration. To take into account the finite transverse
size of the tube was an important task, since in reality a
vortex-forming solenoid is of finite width\footnote{Also, a real
solenoid is of finite length. However, in the case when the width of
the solenoid is much smaller than its length and the motion of a
quantum-mechanical particle is confined to a plane which is
transverse to the solenoid, the effects of the width  prevail over
the effects of the length.}. This task was fulfilled at one time,
see \cite{Olar,Skar}. Although, unlike the case of a singular
vortex, the quantum-mechanical problem in the case of a
finite-thickness vortex is not exactly solvable, a thorough analysis
has been carried out, and, in particular, it has been shown that the
Bohm-Aharonov effect disappears at a sufficiently large thickness of
an impenetrable magnetic vortex  \cite{Olar}.

Returning to quantum field theory, and, appropriately, to the
Casimir-Bohm-Aharonov effect, we note that up to now this effect was
considered for the case of a singular magnetic vortex only
\cite{Sit,Sit1,our1,our2}. Therefore the aim of the present paper is
to make a first step in the study of the dependence on the thickness
of an impenetrable  magnetic vortex. It should be noted that vacuum
polarization effects which are induced by magnetic fluxes of finite
thickness were considered by different authors, see
\cite{Fry,Dunne,BordagKr,Lan,Gra}. However, these authors are
concerned with the case when there is no boundary at all and the
region of the flux is penetrable for the quantized matter fields;
therefore, the obtained results  have no relation neither to the
Casimir, nor to the Bohm-Aharonov effects. In the present paper we
shall show that, similarly to the Bohm-Aharonov effect, the
Casimir-Bohm-Aharonov effect disappears at a sufficiently large
thickness of the vortex. The second-quantized matter will be
represented by the charged massive scalar field.

In the next Section a general definition of the vacuum energy
density for the quantized scalar field is reviewed and a starting
expression for its renormalized  value  is given. In Section 3 this
value is calculated numerically in the case of $(2+1)$-dimensional
space-time. Finally, the results are summarized and discussed in
Section 4.

\section{Vacuum energy density}

The operator of the quantized charged scalar field is represented in
the form
\begin{equation}\label{a11}
 \Psi(x^0,\bi{x})=\sum\hspace{-1.4em}\int\limits_{\lambda}\frac1{\sqrt{2E_{\lambda}}}\left[e^{-{\rm i}E_{\lambda}x^0}\psi_{\lambda}(\bi{x})\,a_{\lambda}+
  e^{{\rm i}E_{\lambda}x^0}\psi_{-\lambda}(\bi{x})\,b^\dag_{\lambda}\right],
\end{equation}
where $a^\dag_\lambda$ and $a_\lambda$ ($b^\dag_\lambda$ and
$b_\lambda$) are the scalar particle (antiparticle) creation and
destruction operators satisfying commutation relations; wave
functions $\psi_\lambda(\textbf{x})$ form a complete set of
solutions to the stationary Klein-Gordon equation
\begin{equation}\label{a12}
 \left(-{\mbox{\boldmath $\nabla$}}^2  + m^2\right)  \psi_\lambda(\bi{x})=E^2_\lambda\psi(\bi{x}),
\end{equation}
$\mbox{\boldmath $\nabla$}$ is the covariant derivative in an
external (background) field and $m$ is the mass of the scalar
particle; $\lambda$ is the set of parameters (quantum numbers)
specifying the state; $E_\lambda=E_{-\lambda}>0$ is the energy of
the state; symbol
  $\sum\hspace{-1em}\int\limits_\lambda$ denotes summation over discrete and
  integration (with a certain measure) over continuous values of
  $\lambda$.

We are considering the static background in the form of the
cylindrically symmetric magnetic vortex of finite thickness, hence
the covariant derivative is $\mbox{\boldmath
$\nabla$}=\mbox{\boldmath $\partial$}-{\rm i}e \bi{V}$ with the
vector potential possessing only one nonvanishing component given by
\begin{equation}\label{3}
V_\varphi=\Phi/2\pi
\end{equation}
outside the vortex; here $\Phi$ is the vortex flux and $\varphi$ is
the angle in the polar $(r,\varphi)$ coordinates on a plane which is
transverse to the vortex. The Dirichlet boundary condition on the
edge $(r=r_0)$ of the vortex is imposed on the scalar field:
\begin{equation}\label{4}
\left.\psi_\lambda\right|_{r=r_0}=0,
\end{equation}
i.e. quantum matter is assumed to be perfectly reflected from the
thence impenetrable vortex. Provided the orthonormalization
condition is satisfied,
\begin{equation}\label{5}
\int d^3x
\psi_\lambda^*\psi_{\lambda'}=\langle\lambda|\lambda'\rangle,
\end{equation}
the solution to \eref{a12} and \eref{4} in the case of the
impenetrable magnetic vortex of thickness $2r_0$ takes form
\begin{equation}\label{a22}
\eqalign{\psi_{knk_z\,}(\bi{x})=(2\pi)^{-1}e^{{\rm i}k_zz}e^{{\rm
i}n\varphi}\beta_{n}(kr_0)\times \cr \times
  [Y_{|n-e\Phi/2\pi|}(kr_0)J_{|n-e\Phi/2\pi|}(kr)-J_{|n-e\Phi/2\pi|}(kr_0)Y_{|n-e\Phi/2\pi|}(kr)],}
\end{equation}
where $z$ is the coordinate along the vortex,
\begin{equation}\label{7}
\beta_n(kr_0)=\left[Y^2_{|n-e\Phi/2\pi|}(kr_0)+J^2_{|n-e\Phi/2\pi|}(kr_0)\right]^{-1/2}\!\!,
\end{equation}
and $0<k<\infty$,  $-\infty<k_z<\infty$, $n\in \mathbb{Z}$
($\mathbb{Z}$ is the set of integer numbers); $J_\mu(u)$ and
$Y_\mu(u)$ are the Bessel functions of order $\mu$ of the first and
second kinds. It should be noted that the vortex can be obviously
generalized to $d$-dimensional space by adding extra $d-3$
longitudinal coordinates to $z$; then factor $(2\pi)^{-1}e^{{\rm
i}k_zz}$ is changed to $(2\pi)^{\frac{1-d}2}e^{{\rm
i}\bi{k}_z\bi{z}}$, where $\bi{z}$ is the $(d-2)$-dimensional vector
which is orthogonal to the $(r,\varphi)$-plane in $d$-dimensional
space.

In general, the vacuum energy density is determined as the vacuum
expectation value of the time-time component of the energy-momentum
tensor, that is given formally by expression
\begin{equation}\label{a14}
\varepsilon=\langle {\rm
vac}|\left(\partial_0\Psi^+\partial_0\Psi+\partial_0\Psi\partial_0\Psi^+\right)|{\rm
vac}\rangle
=\sum\hspace{-1.4em}\int\limits_{\lambda}E_\lambda\psi^*_\lambda(\bi{x})\,\psi_\lambda(\bi{x}),
\end{equation}
which is ill-defined, suffering from the ultraviolet divergencies:
the momentum integral corresponding to the last expression in
\eref{a14} diverges as $p^{d+1}$ for $p\rightarrow\infty$. The
well-defined quantity is obtained with the use of regularization and
then renormalization procedures (see, e.g., \cite{Most}). As to
regularization, one employs conventionally either heat-kernel or
zeta-function methods (see, e.g., \cite{Eli}). As to
renormalization, it has been shown \cite{Sit2} that, for a specific
configuration of a vortex through the excluded region, it suffices,
irrespective of the number of spatial dimensions, to perform one
subtraction, namely to subtract the contribution corresponding to
the absence of the vortex. This fact owes to the symmetry in the
problem, being of rather general nature. It is consistent, for
instance, with a more recent result obtained in a quite different
setup in paper \cite{Wirzba}, where the Casimir energy per unit
length for $n$ non-overlapping parallel cylinders of infinite length
in three-dimensional space is shown to be directly related (without
the need of an extra subtraction or an extra counter-term) to the
Casimir energy for $n$ non-overlapping discs in two-dimensional
space.

Thus, the renormalized vacuum energy density in the case of the
finite-thickness vortex takes form
\begin{equation}\label{c2}
\fl\varepsilon_{ren}= (2\pi)^{1-d}\int
d^{d-2}k_z\int\limits_0^\infty
  dk\,k\left({\bi{k}_z}^2+k^2+m^2\right)^{1/2}\left[S(kr,kr_0)-S(kr,kr_0)|_{\Phi=0}\right],
\end{equation}
where, in view of \eref{a22},
\begin{equation}\label{a29a}
\eqalign{\fl S(kr,kr_0)=\sum_{n\in\mathbb
 Z}\beta_n^2(kr_0)\times\cr
 \times\left[Y_{|n-e\Phi/2\pi|}(kr_0)J_{|n-e\Phi/2\pi|}(kr)-J_{|n-e\Phi/2\pi|}(kr_0)Y_{|n-e\Phi/2\pi|}(kr)\right]^2.}
\end{equation}
Owing to the infinite range of summation, the last expression is
periodic in flux $\Phi$ with period equal to $2\pi e^{-1}$, i.e. it
depends on quantity
\begin{equation}\label{a29a1}
    F=\frac{e\Phi}{2\pi}-\left[\!\left[\frac{e\Phi}{2\pi}\right]\!\right],
\end{equation}
where $[[u]]$ is the integer part of quantity $u$ (i.e. the integer
which is less than or equal to $u$).

Let us rewrite \eref{a29a} in the form
\begin{equation}\label{a29b}
S(kr,kr_0)=S_{0}(kr)+S_{1}(kr,kr_0),
\end{equation}
where $S_{0}(kr)$ corresponds to the appropriate series in the case
of the vacuum polarization by  a singular magnetic vortex
\cite{Sit1,our1,our2}:
\begin{equation}\label{a29b1}
\fl S_0(kr)= \sum_{n=0}^\infty\left[
J^2_{n+F}(kr)+J^2_{n+1-F}(kr)\right] =\int\limits_0^{kr}\!
 d\tau\left[J_F(\tau)J_{-1+F}(\tau)+J_{-F}(\tau)J_{1-F}(\tau)\right],
\end{equation}
and $S_1(kr,kr_0)$ is a correction term due to the finite thickness
of a vortex:
 \begin{equation}\label{a29c}
\eqalign{\fl S_1(kr,kr_0) = 2\sum_{n=0}^\infty
\left[J_{n+F}(kr_0)Y_{n+F}(kr)\frac{J_{n+F}(kr_0)Y_{n+F}(kr)-Y_{n+F}(kr_0)J_{n+F}(kr)}{J_{n+F}^2(kr_0)+Y_{n+F}^2(kr_0)}\right.+\cr\fl
 +\left.
J_{n+1-F}(kr_0)Y_{n+1-F}(kr)\frac{J_{n+1-F}(kr_0)Y_{n+1-F}(kr)-Y_{n+1-F}(kr_0)J_{n+1-F}(kr)}{J_{n+1-F}^2(kr_0)+Y_{n+1-F}^2(kr_0)}\right]-\cr\fl
-\sum_{n=0}^\infty\left[J_{n+F}^2(kr_0)\frac{J_{n+F}^2(kr)+Y_{n+F}^2(kr)}{J_{n+F}^2(kr_0)+Y_{n+F}^2(kr_0)}
+J_{n+1-F}^2(kr_0)\frac{J_{n+1-F}^2(kr)+Y_{n+1-F}^2(kr)}{J_{n+1-F}^2(kr_0)+Y_{n+1-F}^2(kr_0)}\right].}
\end{equation}

 In the absence of the magnetic flux in the tube we have
 \begin{equation}\label{c1a}
S(kr,kr_0)|_{\Phi=0}=\tilde S_0+\tilde S_1(kr,kr_0),
\end{equation}
where
\begin{equation}\label{c1b}
\tilde S_0=J^2_{0}(kr)+ 2\sum_{n=1}^\infty J^2_{n}(kr)=1,
\end{equation}
and a correction term due to the finite thickness of an empty tube:
\begin{equation}\label{c1c}
\eqalign{\fl  \tilde S_1(kr,kr_0)=
2\left[J_{0}(kr_0)Y_{0}(kr)\frac{J_{0}(kr_0)Y_{0}(kr)-Y_{0}(kr_0)J_{0}(kr)}{J_{0}^2(kr_0)+Y_{0}^2(kr_0)}\right.+\cr
+\left. 2\sum_{n=1}^\infty
J_{n}(kr_0)Y_{n}(kr)\frac{J_{n}(kr_0)Y_{n}(kr)-Y_{n}(kr_0)J_{n}(kr)}{J_{n}^2(kr_0)+Y_{n}^2(kr_0)}\right]-\cr
-\left[J_{0}^2(kr_0)\frac{J_{0}^2(kr)+Y_{0}^2(kr)}{J_{0}^2(kr_0)+Y_{0}^2(kr_0)}
+2\sum_{n=1}^\infty
J_{n}^2(kr_0)\frac{J_{n}^2(kr)+Y_{n}^2(kr)}{J_{n}^2(kr_0)+Y_{n}^2(kr_0)}\right].}
\end{equation}

Thus, vacuum energy density \eref{c2} depends on $F$ \eref{a29a1},
i.e. it is periodic in flux $\Phi$ with a period equal to $2\pi
e^{-1}$. Moreover, relation \eref{c2} is symmetric under
substitution $F\rightarrow1-F$, vanishing at $F\rightarrow0$
$(F\rightarrow1)$ and, perhaps, attaining its maximal value at
$F=1/2$\footnote{At least, this is certainly true in the case of the
singular vortex both for the Bohm-Aharonov \cite{Aha} and the
Casimir-Bohm-Aharonov \cite{Sit, Sit1, our1,our2} effects.}.
Relations \eref{a29b1} and \eref{a29c} are simplified at $F=1/2$:
\begin{equation}\label{18}
 S_0(kr)|_{\Phi=\pi
 e^{-1}}=\frac2\pi\int\limits_{0}^{2kr}\frac{d\,\tau}{\tau}\sin\tau,
\end{equation}
and
 \begin{equation}\label{19}
\eqalign{\fl S_1(kr,kr_0)|_{\Phi=\pi
 e^{-1}} =\cr\fl  =\!2\!\sum_{n=0}^\infty
\frac{J_{n+\frac12}^2(kr_0)\!\left[Y_{n+\frac12}^2(kr)\!-\!J_{n+\frac12}^2(kr)\right]\!-\!2J_{n+\frac12}(kr_0)Y_{n+\frac12}(kr_0)J_{n+\frac12}(kr)Y_{n+\frac12}(kr)}
{J_{n+\frac12}^2(kr_0)+Y_{n+\frac12}^2(kr_0)}.}
\end{equation}

Since it is hardly possible to evaluate sums in \eref{a29c} and
\eref{c1c} analytically, our further analysis will employ numerical
calculation. In the following we restrict ourselves to the case of
$F=1/2$ and $d=2$, when the expression for the vacuum energy density
takes form
\begin{equation}\label{c3a}
\varepsilon_{ren}=\frac{1}{2\pi}\int\limits_0^\infty
dk\,k\left(k^2+m^2\right)^{1/2}G(kr,kr_0),
\end{equation}
where
\begin{equation}\label{21}
G(kr,kr_0)=S(kr,kr_0)|_{\Phi=\pi e^{-1}}-S(kr,kr_0)|_{\Phi=0}.
\end{equation}

\section{Numerical evaluation of the vacuum energy density}

We rewrite \eref{c3a} in the dimensionless form
\begin{equation}\label{c3}
r^3\varepsilon_{ren}=\frac{1}{2\pi}\int\limits_0^\infty
dz\,z\sqrt{z^2+\left(\frac{mr_0}\lambda\right)^2} G(z,\lambda z),
\end{equation}
where $\lambda=r_0/r$, $\lambda\in[0,1]$. Let us point out some
analytical properties of the integrand function in \eref{c3}: it
vanishes at the edge of the vortex
\begin{equation}\label{23}
\lim_{\lambda\rightarrow1}G(z,\lambda z)=0;
\end{equation}
at large distances from the vortex the case of a singular vortex is
recovered
\begin{equation}\label{24}
\lim_{\lambda\rightarrow0}G(z,\lambda z)=S_0(z)|_{\Phi=\pi
e^{-1}}-\tilde S_0;
\end{equation}
at small values of $z$ one gets
\begin{equation}\label{c6}
G(z,\lambda z)|_{z\rightarrow0}=-[\ln(\lambda)/\ln(\lambda z)]^2.
\end{equation}

Numerical analysis indicates that in the calculation of function
$G(z,\lambda z)$  one can use  series in  \eref{c1c} and \eref{19}
with finite limits, namely for calculation $G(z,\lambda z)$ at point
$z=z'$  it is enough to cut off the summation limits by
$n=[[z'+30]]$. In this case the relative error is
\begin{equation}\label{c4a}
\fl\left|\frac{G(z,\lambda z)|_{n\in(0,[[z+30]])}-G(z,\lambda
z)}{G(z,\lambda
z)}\right|<\delta(\lambda),\quad\delta(\lambda)<10^{-17},\quad\lambda\in[1/10,9/10].
\end{equation}
It can be shown that the envelope of $G(z,\lambda z)$ is
exponentially decreasing function at large $z$, see Fig.1. So, for
the finite-thickness magnetic vortex  we can compute values of
dimensionless quantity $r^3\varepsilon_{ren}$ \eref{c3} for
different (not very small) values of $\lambda$.  To do this, we have
to be able to perform integration in (\ref{c3}) with high precision.
We make it in a following way.

\begin{figure}[h]
\begin{center}
\includegraphics[width=90mm]{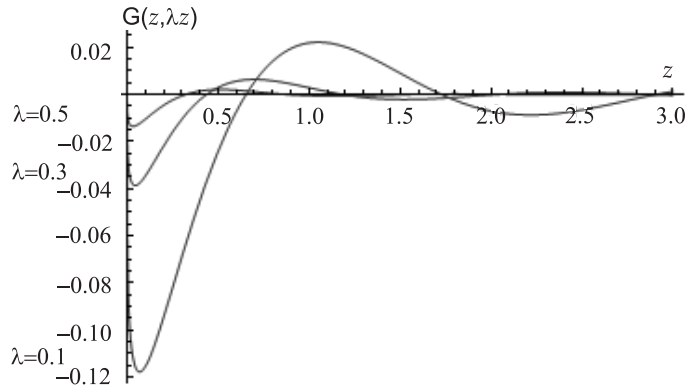}
\end{center}
\vspace{-1em} \caption{Behavior of $G(z,\lambda z)$ at different
values of $\lambda$.}
\end{figure}

As one can see from Fig.2, the function $G(z,\lambda z)$ is negative
from $z=0$ to the first function root at $z=z_1$ ($z_1\neq0$). So,
the appropriate integral in \eref{c3} is negative. The subsequent
roots are denoted by $z_2$, $z_3$, etc. Because of decreasing
character of the envelope function the integral from $z_1$ to $z_3$
will be positive. It is useful to define a period of function
$G(z,\lambda z)$ as an interval between two next to neighboring
roots, i.e. from $z_1$ to $z_3$, from $z_3$ to $z_5$, and so on.
Then the full integral in \eref{c3} will be a sum of the negative
integral from $z=0$ to $z=z_1$ and a multitude of positive integrals
over subsequent periods. In the case of sufficiently small
transverse size of the tube ($mr_0<0.1$) the integrals over some
finite number of first periods may be negative but thereupon they
become and remain positive also.

For small $z$ ($ z\lesssim20$) we make a direct integra\-tion of
function $G(z,\lambda z)$ over periods  using 25 digits of precision
in internal com\-pu\-ta\-tions.

\begin{figure}[b]
\begin{center}
\includegraphics[width=90mm]{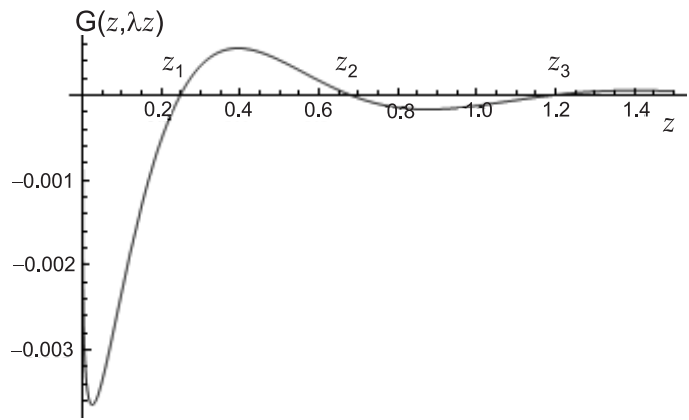}
\end{center}
\vspace{-1em}\caption{The location of roots of $G(z,\lambda z)$  at
$\lambda=0.7$.}
\end{figure}

For large $z$ we make integration for each period separately. To do
it we create a table of values of  function $G(z,\lambda z)$ for a
separate period  and replace this function by more simple function
in the form
\begin{equation}\label{c8}
G_{int}(z,\lambda z)=a\frac{e^{-b
z}}{z^c}\frac{A_q(z^2)}{B_q(z^2)}\sin(k z+j\ln z+\phi_0),
\end{equation}
where sine function ensures that roots of $G_{int}(z,\lambda z)$
coincide with roots of $G(z,\lambda z)$; $A_{q}(y)$ and $B_{q}(y)$
are $q$-degree polynomials, $q$ can be 3,\,4 or 5; all unknown
parameters can be found from an interpolation procedure. We allow a
relative error of interpolation to be
\begin{equation}\label{c9}
\left|\frac{G_{int}(z,\lambda z)-G(z,\lambda z)}{G(z,\lambda
z)}\right|<10^{-8}
\end{equation}
for each period. The function $G_{int}(z,\lambda z)$ can be
immediately integrated with the required accuracy. In this way we
made integration up to $z\simeq100/\lambda$ with absolute accuracy
up to $10^{-17}$.

\begin{figure}[t]
\begin{center}
\includegraphics[width=90mm]{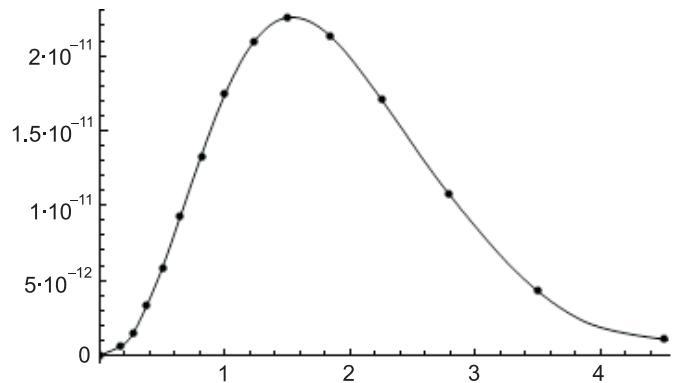}
\end{center}
\vspace{-1em}\caption{$r^3\varepsilon_{ren}$ at $mr_0=3/2$ as a
function of $x$.}
\end{figure}

With the help of the above procedure  we obtain a table of
contributions from integration over each period, extrapolate this
table to infinity, and after that we find the full integral in
\eref{c3} as a sum of the negative integral over first period(s), a
multitude of positive integrals over periods up to
$z\simeq100/\lambda$ and an interpolation term. The absolute
accuracy of the obtained result is $10^{-13}$. It should be noted
that nearly 99 \% of the integral value in \eref{c3} is obtained by
direct calculation and only nearly one percent is the contribution
from the interpolation.

\begin{figure}[b]
\begin{center}
\includegraphics[width=90mm]{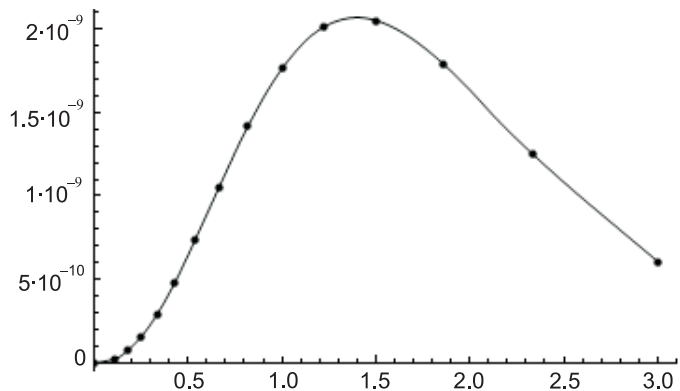}
\end{center}
\vspace{-1em}\caption{$r^3\varepsilon_{ren}$ at $mr_0=1$ as a
function of $x$.}
\end{figure}

Dimensionless quantity  $r^3\varepsilon_{ren}$ \eref{c3} is a
function of two dimensionless parameters, $mr_0$ and $mr$. Using the
above described procedure, we calculate  $r^3\varepsilon_{ren}$ at
several values of $mr_0$ as a function of dimensionless distance
from the edge of the vortex, $x=m(r-r_0)$, in the range $0<x<3mr_0$.
Further increase of the distance from the vortex results in a
significant increment of computational time, because there the
envelope of $G(z,\lambda z)$ fails to be a sufficiently decreasing
function as it is at smaller distances. The results of our numerical
calculations are presented in Figs.3--7, where
$r^3\varepsilon_{ren}$ is along the ordinate axis and $x$ is along
the abscissa axis; solid lines are interpolating the dots that have
been calculated.

\begin{figure}[t]
\begin{center}
\includegraphics[width=90mm]{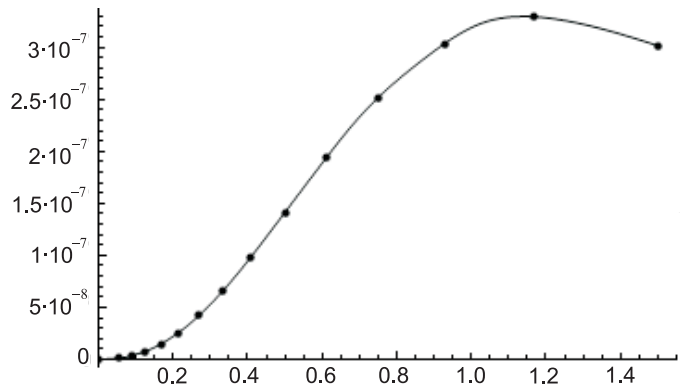}
\end{center}
\vspace{-1em}\caption{$r^3\varepsilon_{ren}$ at $mr_0=1/2$ as a
function of $x$.}
\end{figure}

The typical behaviour of $r^3\varepsilon_{ren}$ is clearly
illustrated in cases $mr_0\geq1$ by Fig.3 and Fig.4 The vacuum
energy density is zero at the edge of the vortex (at $x=0$), starts
increasing by some power law $x^\alpha$ with $\alpha>1$, reaches
maximum at $x\sim1$, and decreases at larger distances to zero
(probably exponentially as $e^{-x}$). However, as $mr_0$ decreases,
the available range of $x$ is shrunk due to above mentioned
restriction $x<3mr_0$. In the case of $mr_0=1/2$ a maximum at
$x\sim1$ is clearly seen (Fig.5), and a following decrease to zero
may be anticipated. In the cases of $mr_0=10^{-1}$ (Fig.6) and
$mr_0=10^{-4}$ (Fig.7) one may suppose that there will be a maximum
at $x\sim1$ and a following decrease to zero. But, one can be sure
for certain from Figs.3--7 that the vacuum energy density decreases
to zero as $x^\alpha$ with $\alpha>1$ at $x\rightarrow0$.

\begin{figure}[h]
\begin{center}
\includegraphics[width=90mm]{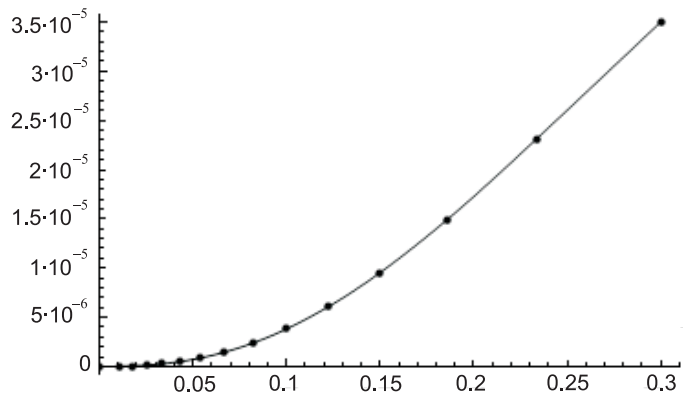}
\end{center}
\vspace{-1em}\caption{$r^3\varepsilon_{ren}$ at $mr_0=10^{-1}$ as a
function of $x$.}
\end{figure}

As to values of the vacuum energy density, they are rapidly
decreasing as parameter $mr_0$ increases and becomes more than
unity. Namely, the maximal values of $r^3\varepsilon_{ren}$ are:
$3.3\cdot10^{-7}$ at $mr_0=1/2$ (Fig.5), $2.1\cdot10^{-9}$ at
$mr_0=1$ (Fig.4) and $2.2\cdot10^{-11}$ at $mr_0=3/2$ (Fig.3). These
should be compared with much larger values which are already
attained below maxima in Fig.6 and Fig.7: $3.5\cdot10^{-5}$ at
$mr_0=10^{-1}$ and $7\cdot10^{-5}$ at $mr_0=10^{-4}$. It should be
noted that, in the case of the singular vortex ($mr_0=0$), the
maximal value of $r^3\varepsilon_{ren}$ is
$(12\pi^2)^{-1}\approx8.5\cdot10^{-3}$ \cite{Sit1}; the appropriate
plot of $r^3\varepsilon_{ren}$ as a function of $mr$ is taken from
\cite{our2} and is presented in Fig.8. Thus, one may suppose that,
in the case of the vortex with thickness in the range
$0<mr_0<10^{-4}$, the maximal value of $r^3\varepsilon_{ren}$ will
be somewhere in the range $10^{-4}\div10^{-3}$. For more clarity,
the results of Fig.6 and Fig.7 are plotted as functions of variable
$r/r_0=\lambda^{-1}$ in Fig.9. Note that, as $mr_0$ falls by three
orders from $10^{-1}$ to $10^{-4}$, quantity $r^3\varepsilon_{ren}$
changes  by factor $2$ only. This should be compared with
$r^3\varepsilon_{ren}$ at $m=0$, which is plotted as a function of
$r/r_0$ in Fig.10; the latter plot coincides actually with that
corresponding to $mr_0=10^{-4}$ in Fig.9. It should be noted also
that at sufficiently small distances from the vortex edge (at
$r-r_0\ll m^{-1}$) the behaviour of $r^3\varepsilon_{ren}$ coincides
with that in the $m=0$ case.

 \begin{figure}[t]
\begin{center}
\includegraphics[width=90mm]{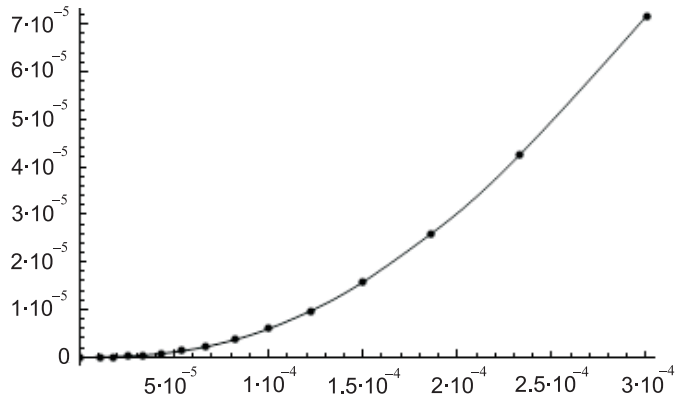}
\end{center}
\vspace{-1em}\caption{$r^3\varepsilon_{ren}$ at $mr_0=10^{-4}$ as a
function of $x$.}
\end{figure}

\begin{figure}[b]
\begin{center}
\includegraphics[width=90mm]{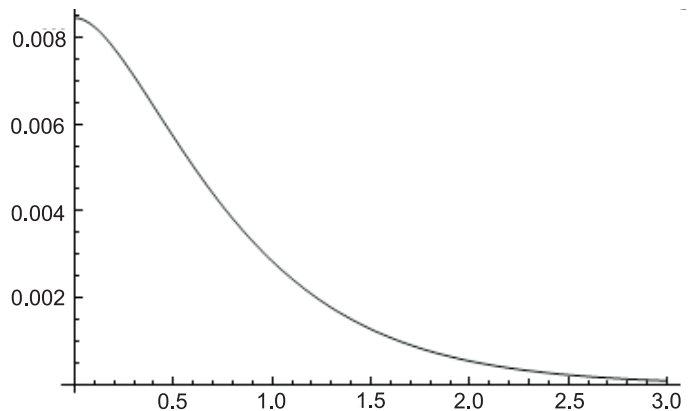}
\end{center}
\vspace{-1em}\caption{$r^3\varepsilon_{ren}$ at $r_0=0$ as a
function of $mr$.}
\end{figure}

\section{Discussion of the results}

We have studied the influence of finite thickness of the
impenetrable magnetic vortex on the vacuum polarization of the
quantized charged massive scalar field. Since units $c=\hbar=1$ are
used, the London flux value is $2\pi e^{-1}$, and we show that
induced vacuum energy density \eref{c2} is periodic in the value of
vortex flux $\Phi$, vanishing at integer multiples of the London
flux value (at $\Phi=2\pi n e^{-1}$) and being presumably maximal at
half of the London flux value (at $\Phi=\pi (2n+1) e^{-1}$). If the
vortex thickness decreases, $r_0\rightarrow0$, or a distance from
the vortex increases, $r-r_0\rightarrow\infty$, then the
contribution of
 $S_1(kr,kr_0)$ \eref{a29c} and $\tilde S_1(kr,kr_0)$ \eref{c1c} to
\eref{c2} tends smoothly to zero, and the vacuum energy density
converges with that induced by the singular magnetic vortex.

 \begin{figure}[b]
\begin{center}
\includegraphics[width=90mm]{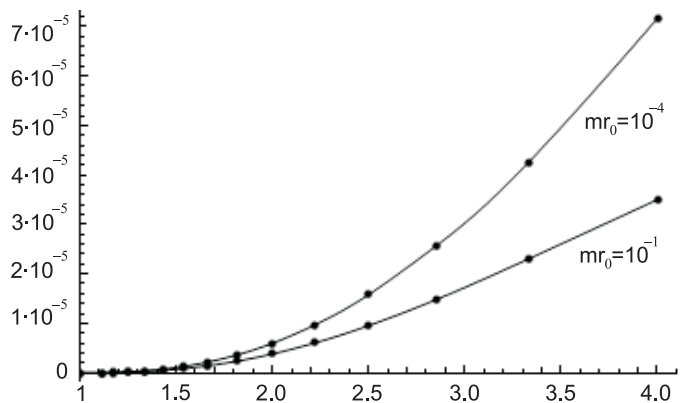}
\end{center}
\vspace{-1em}\caption{$r^3\varepsilon_{ren}$ at the smallest values
of $mr_0$ as a function of $r/r_0$.}
\end{figure}

Our numerical analysis of the vortex thickness effects has been
carried out for the case of the vortex flux equal to half of the
London flux value; the quantized scalar field is confined to a plane
which is orthogonal to the vortex. As follows from this analysis,
the vacuum polarization actually disappears, when the transverse
size of the vortex $(r_0)$ exceeds the Compton wavelength of the
scalar particle $(m^{-1})$: the maximal value of the induced vacuum
energy density falls by two orders from $2.6\cdot10^{-10}m^3$ to
$1.4\cdot10^{-12}m^3$ as $mr_0$ increases from $1$ to $3/2$.

This result should be compared with the influence of the vortex
thickness on the conventional Bohm-Aharonov effect. In the framework
of first-quantized theory, one considers elastic scattering of a
quantum-mechanical charged particle on the impenetrable magnetic
vortex of thickness $2r_0$. The incident wave is characterized by
momentum $p$, so the dimensionless parameter of the problem is
$pr_0$. In the long-wavelength limit, $pr_0\rightarrow0$, scattering
converges with scattering on the singular magnetic vortex
\cite{Skar}. Since the short-wavelength limit,
$pr_0\rightarrow\infty$, corresponds to the case when
quasi-classical approximation is applicable, one would anticipate
that the purely quantum effect, as is the Bohm-Aharonov one,
disappears in this limit. As has been shown in \cite{Olar}, this
anticipation is indeed confirmed, and scattering in the
$pr_0\rightarrow\infty$ limit
 converges with scattering of a classical
point particle on the impenetrable tube, being independent of the
enclosed magnetic flux.

In the framework of second-quantized theory, one considers the
vacuum polarization in the background of the impenetrable magnetic
vortex. The appropriate dimensionless parameter is $mr_0$, and, as
we have shown in the present paper, the Casimir-Bohm-Aharonov effect
disappears in the $mr_0\rightarrow\infty$ limit, becoming actually
negligible at $mr_0>3/2$.

In the case of the singular magnetic vortex, the induced vacuum
energy density diverges at the location of the vortex
\cite{Sit,Sit1,our1,our2}. As it has been shown in the present
paper, this divergence is unphysical, disappearing when thickness of
the impenetrable magnetic vortex is taken into account: under the
Dirichlet condition for the quantized field \eref{4}, the induced
vacuum energy density is vanishing as $(r-r_0)^\alpha$ with
$\alpha>1$ at the edge of the vortex. Therefore, the vacuum energy
which is induced on the whole transverse plane,
\begin{equation}\label{29}
E_{ren}=2\pi\int\limits_{r_0}^\infty dr\,r\varepsilon_{ren},
\end{equation}
is finite, contrary to the case of the singular vortex when it is
infinite. Although we are unaware  of the value of $E_{ren}$, the
maximal value of $\varepsilon_{ren}$ is estimated to be somewhat of
the order of $10^{-3}m^3$ if $mr_0<10^{-4}$.

A brief discussion of polarization of the vacuum of the quantized
massless scalar field is in order.  In this case, the induced vacuum
energy density is zero at the edge of the vortex, starts increasing
as $(r-r_0)^\alpha$ with $\alpha>1$ (see Fig.10), reaches its
maximum   and then decreases with asymptotics $(12\pi^2r^3)^{-1}$
\cite{Sit1,our1,our2}. Induced vacuum energy \eref{29} is finite in
this case also.

 \begin{figure}[b]
\begin{center}
\includegraphics[width=90mm]{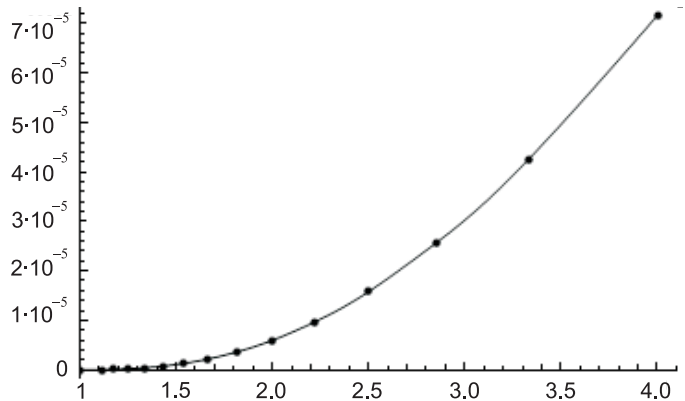}
\end{center}
\vspace{-1em}\caption{$r^3\varepsilon_{ren}$ at  $m=0$ as a function
of $r/r_0$.}
\end{figure}

The finite-thickness vortex can be formed as a topological defect
appearing after a phase transition with spontaneous breakdown of the
gauge symmetry \cite{Abr}. Such a structure under the name of a
cosmic string \cite{vilenkin,hind} is currently discussed in various
contexts in cosmology and astrophysics, see, e.g.,
\cite{pol,sazhin}. The cosmic string is characterized by flux $2\pi
e^{-1}_{{\rm H}}$, where $e_{\rm H}$ is the coupling constant of the
Higgs scalar field to the string-forming gauge field; the transverse
size of the string is of the order of correlation length $m_{\rm
H}^{-1}$, where $m_{\rm H}$ is the mass of the Higgs scalar field.
Then, as it follows from our consideration in the present paper, the
cosmic string can polarize the vacuum of quantum matter only in the
case when the mass of the matter field is much less than that of the
Higgs field, $m\ll m_{\rm H}$. For instance, the cosmic string which
is formed at the grand unification scale can polarize the vacuum of
the electroweak theory, whereas the would-be cosmic string
corresponding to the electroweak symmetry breaking has no impact on
the vacuum of quantum matter at the grand unification scale.

\ack

Yu A S acknowledges the support from the State Foundation for
Fundamental Research under project F28.2/083 "Application of the
string theory and the field theory methods to the study of nonlinear
phenomena in low-dimensional systems" and from the National Academy
of Science of Ukraine under project 10/07-N "Study of physical
properties of nanomaterials for electronics, photonics, spintronics
and information technologies".

\section*{References}

%%%%%%%%%%%%%%%%%        BIBLIOGRAPHY

\end{document}